\documentclass[aps,prd,twocolumn,superscriptaddress,floatfix,nofootinbib]{revtex4-2}
\pdfoutput=1
\usepackage[T1]{fontenc}
\usepackage[english]{babel}
\usepackage{amsmath}
\usepackage{amssymb}
\usepackage{amsthm}
\usepackage{amsfonts}
\usepackage{graphicx}
\usepackage{enumitem}
\usepackage{bm}
\usepackage{rotating}
\usepackage[dvipsnames]{xcolor}
\definecolor{myblue}{HTML}{332288}
\usepackage[colorlinks]{hyperref}
 \hypersetup{
     colorlinks=true,
     linkcolor=myblue,
     filecolor=myblue,
     citecolor=myblue,
     urlcolor=myblue,
     }
\usepackage{pbox}

\usepackage{array}
\usepackage{physics}
\usepackage{mathtools}
\usepackage{leftidx}
\usepackage{lmodern}
\usepackage[normalem]{ulem}
\usepackage{braket}
\usepackage{tensor}
\usepackage{mathrsfs}
\usepackage{float}
\usepackage{dsfont}
\usepackage[margin=2cm]{geometry}
\usepackage[title]{appendix}
\usepackage{tcolorbox}
\usepackage{tocloft}
\usepackage{titlesec}
\usepackage[caption=false]{subfig}

\usepackage{tikz}
\usetikzlibrary{arrows.meta,decorations.pathmorphing,math}

\renewcommand{\Re}{\mathrm{Re}}
\renewcommand{\Im}{\mathrm{Im}}

\renewcommand{\tilde}{\widetilde}

\newcommand{\ii}{\mathsf{i}}

\newcommand{\iu}{\mathrm{i}\mkern1mu}

\newcommand{\sx}{\mathsf{x}}

\newcommand{\bx}{{\bm{x}}}

\newcommand{\bk}{{\bm{k}}}

\newcommand{\rod}{\rho_{\textsc{d}}^{(0)}}

\newcommand{\ra}{\rho_{\textsc{a}}}
\newcommand{\rb}{\rho_{\textsc{b}}}
\newcommand{\rof}{\rho_{\phi}^{(0)}}
\newcommand{\rd}{\rho_{\textsc{d}}}
\newcommand{\roo}{\rho^{(0)}}

\newcommand{\Z}{\mathbb{Z}}

\newcommand{\superprime}{^{\prime}}

\definecolor{darkraspberry}{rgb}{0.53,0.15,0.34}

\begin{document}

\title{Contextuality from the vacuum}

\author{Caroline Lima}
\email{clima@perimeterinstitute.ca}
\affiliation{Department of Physics and Astronomy, University of Waterloo, Waterloo, Ontario, N2L 3G1, Canada}
\affiliation{Institute for Quantum Computing, University of Waterloo, Waterloo, Ontario, N2L 3G1, Canada}
\affiliation{Perimeter Institute for Theoretical Physics, 31 Caroline St N, Waterloo, Ontario, N2L 2Y5, Canada}
\author{María Rosa Preciado-Rivas}%
\email{mrpreciadorivas@uwaterloo.ca}
\affiliation{Institute for Quantum Computing, University of Waterloo, Waterloo, Ontario, N2L 3G1, Canada}
\affiliation{Department of Applied Mathematics, University of Waterloo, Waterloo, Ontario, N2L 3G1, Canada}
\affiliation{Waterloo Centre for Astrophysics, University of Waterloo, Waterloo, Ontario, N2L 3G1, Canada}
\author{Sanchit Srivastava}
\email{sanchit.srivastava@uwaterloo.ca}
\affiliation{Department of Physics and Astronomy, University of Waterloo, Waterloo, Ontario, N2L 3G1, Canada}
\affiliation{Institute for Quantum Computing, University of Waterloo, Waterloo, Ontario, N2L 3G1, Canada}

\date{\today}

\begin{abstract}
Contextuality, a key resource for quantum advantage, describes systems in which the outcome of a measurement is \emph{not} independent of other compatible measurements, in contrast to classical hidden-variable descriptions. We investigate the harvesting of contextuality from the vacuum of a quantum field using Unruh-DeWitt detectors. We show that localized interactions with the field can endow initially  non-contextual detectors with contextuality with respect to Heisenberg-Weyl measurements, as quantified by contextual fraction. The harvested contextuality correlates with the emergence of Wigner function negativity, in agreement with known equivalences between these notions. Our results show that contextuality is a resource that can be extracted directly from the quantum vacuum and establish contextuality harvesting as a fundamental phenomenon in relativistic quantum information.
\end{abstract}

\maketitle

\section{Introduction}\label{sec:intro}
\footnotetext{The order of the authors in this article is alphabetical. All the authors contributed equally to this work.}  

Quantum theory rules out hidden-variable models that assign preexisting, context-independent values to all observables, where a context denotes a set of compatible measurements~\cite{kochen_problem_1967}. In particular, any model in which the outcome depends only on the observable being measured, and not on the other observables measured alongside it, cannot reproduce the statistics of quantum measurements. This failure of context-independence is known as contextuality, which has been the subject of extensive study in the foundations of quantum mechanics~\cite{budroni_kochen-specker_2022}.

Contextuality has also emerged as a quantum resource for enabling tasks such as random number generation, self-testing, and measurement-based quantum computation~\cite{Amaral2019, Howard:2014zwm, delfosse_equivalence_2017, raussendorf_contextuality_2013, hu_self-testing_2023}.
 A growing body of work has demonstrated that contextuality is closely connected to other nonclassical resources. Notably, measures such as Wigner function negativity and magic (or non-stabilizerness) have been shown to provide operational advantages, with contextuality serving as a unifying principle underlying these phenomena \cite{Howard:2014zwm, delfosse_equivalence_2017, spekkens_negativity_2008,schmid_uniqueness_2022,schmid_structure_2024}. This resource perspective positions contextuality as both a fundamental conceptual marker of nonclassicality and a practical enabler of quantum computational power.

In parallel, a well established protocol of relativistic quantum information (RQI) --- a field which addresses questions at the intersection of information theory, quantum mechanics, and relativity --- is entanglement harvesting~\cite{valentini_non-local_1991,reznikEntanglementVacuum2003,Pozas-Kerstjens:2015gta}. In this protocol, two Unruh–DeWitt (UDW) detectors~\cite{Unruh.effect, DeWitt1979} can become entangled with one another, even if they are initially uncorrelated, by interacting locally with the vacuum state of a quantum field. 
In this vein, one could ask what other resources a detector can extract through interactions with a quantum field.
Mutual information~\cite{Pozas-Kerstjens:2015gta}, discord~\cite{Brown:2013kia,Lin:2024roh,Fan:2024gai}, non-local coherence~\cite{Wang:2023dyk}, and steering~\cite{Wu:2024whx} have all been studied in this operational approach.
More recently, using mana, \citeauthor{Nystrom:2024oeq} showed that a qutrit UDW detector initially in a non-magical state can become magical after interacting with a quantum field~\cite{Nystrom:2024oeq}, and \citeauthor{Cepollaro:2024sod} studied  non-stabilizerness by calculating the stabilizer Rényi entropy of two UDW accelerated detectors~\cite{Cepollaro:2024sod}.

In this work, we build upon the RQI operational approach and focus on contextuality. Specifically, we consider a two-qutrit system that is initially noncontextual with respect to Heisenberg–Weyl measurements \cite{gottesman_stabilizer_1997, gross_hudsons_2006}, and investigate how its interaction with a quantum field can give rise to a non-zero contextual fraction. This exploration is significant in three regards: first, it constitutes a new protocol within RQI, expanding the scope of questions the field can answer; second, it proves that detectors interacting with a quantum field can acquire this quantum resource, which underlies Wigner negativity and magic; finally, given the foundational significance of contextuality, this perspective offers new insight into the foundational aspects of the quantum field itself.

\footnotetext{
During the final stages of this work, the authors became aware of similar, independent work by P. LeMaitre, which appeared on arXiv the day before our submission~\cite{lemaitre_harvesting_2025}.}

Our paper is organized as follows. In \autoref{sec:setup} we introduce the physical setup of two qutrit detectors interacting with a quantum field. In \autoref{sec:contfraction} we outline the framework of contextuality and contextual fraction, which we use to quantify contextuality in the joint detector state. In \autoref{sec:Wignernegativity} we evaluate inequality violations for the reduced single-qutrit states, corresponding to negativity of the Gross discrete Wigner function \cite{gross_hudsons_2006} (which we will refer to simply as “Wigner negativity”), and relate these results to known connections between contextuality and magic \cite{Howard:2014zwm}. In \autoref{sec:results} we present our results for contextual fraction and Wigner function negativity. Finally, in \autoref{sec:discussion} we summarize and discuss the implications of our findings. \autoref{app:comparisonSU2HW} provides a discussion about how the internal dynamics of the detector influences the harvested contextual fraction; \autoref{app:integrals} gives the explicit integrals that were numerically solved and used in our plots; and \autoref{app:more_plots} shows results for other set of fixed parameters and details how our results compare to the ones in Ref.~\cite{Nystrom:2024oeq}.
We adopt the mostly plus signature, and denote spacetime events by $\mathsf{x}=(t, \bm x)$. We use $c=\hbar=1$.

\section{Setup}\label{sec:setup}
The essence of the proposed contextuality harvesting protocol is the same as the usual entanglement harvesting one, which relies on the Unruh-DeWitt particle detector model~\cite{Unruh.effect, DeWitt1979}. However, instead of the commonly used qubit UDW detector, we model the detectors interacting with the field as two \emph{qutrits}. We provide a reasoning for this choice at the end of the \autoref{sec:contfraction}, once the relevant background has been introduced.

We start with a massless real scalar quantum field in (1+3)-dimensional flat spacetime\footnote{The generalization to $(1+n)$-dimensional globally hyperbolic curved spacetimes is straightforward.}, which admits a plane-wave mode expansion 
\begin{align}
     \hat{\phi}(\sx) = \int\frac{\dd^3\bk}{ \sqrt{2 (2\pi)^3 |\bk|}}\,\hat{a}_\bk e^{-\iu |\bk| t + \iu \bm{k}\cdot\bx} + \text{h.c.},
\end{align}
where the ladder operators $\hat{a}_\bk^\dagger$ and $\hat{a}_\bk$ satisfy the canonical commutation relations, $[\hat{a}_\bk,{\hat{a}^{\dagger}}_{\bk'}]=\delta^{(3)}(\bk-\bk')$.\footnote{Hereafter, we drop the hats as it is clear from the context that we are referring to quantized observables.}

We then take two qutrits (hereafter called detectors $a$ and $b$), such that the free Hamiltonian for each of them is
\begin{equation}\label{eq:freeHamiltonian}
    H_{D,i} = \Omega_{i,0}\ketbra{0}{0}_i+ \Omega_{i,1} \ketbra{1}{1}_i + \Omega_{i,2} \ketbra{2}{2}_i,
\end{equation}
where $i\in\{a,b\}$ and $\left\{ \ket{0}_i, \ket{1}_i, \ket{2}_i\right\}$ is an orthonormal basis. For simplicity, we set the detectors to be comoving with respect to the quantization frame, $\mathsf{x}=(t,\bm{x})$. In the Dirac picture, the Hamiltonian density capturing the interaction of the two detectors with the quantum field is
\begin{subequations}\label{eq:interaction_h}
    \begin{align}
        h_I(\mathsf{x}) &= \sum_i \lambda_i \Lambda_i(\mathsf{x}) O_i(t)\otimes \phi(\sx),\\
        O_i(t) &= e^{\iu H_{D,i} t} O_i e^{-\iu H_{D,i} t},
    \end{align}
\end{subequations}
where $\lambda_i$ is the coupling constant between detector $i$ and the field,  $\Lambda_i(\sx)$ is a smearing function that controls the region of spacetime where detector $i$ interacts locally with the field, and $O_i$ is an observable of detector $i$. 
$\Lambda_i(\sx)$ is defined with respect to $\sx_i=(t_i,\bm{x}_i)$, which is the trajectory followed by the $i$-th detector.
Throughout, we consider the detectors to be identical, with $\lambda_a=\lambda_b \eqqcolon \lambda$.

Initializing the detectors in an uncorrelated state and the field in its vacuum state $\ket{0}$ at $t\to -\infty$, the final state is evaluated at $t \to \infty$. The time evolution operator is thus
\begin{align}
    U &= \mathcal{T}_t \exp\left[-\ii\int \dd \mathsf{x}\, h_I(\mathsf{x})\right]\,,
\end{align}
where $\mathcal{T}_t$ is the time ordering operator with respect to $t$.
For a small coupling constant $\lambda$, the time evolution operator can be expanded into the Dyson series
\begin{subequations}
\begin{align}
    U &= \openone + U^{(1)} + U^{(2)} + \mathcal{O}(\lambda^3)\,, \\ 
   U^{(1)} &= -\ii\int \dd \mathsf{x}\,h_I(\mathsf{x})\,,\\
    U^{(2)} &= -\int \dd \mathsf{x}\int \dd \mathsf{x}'\,\Theta(t-t')h_I(\mathsf{x})h_I(\mathsf{x}')\,,
\end{align}
\end{subequations}
where $U^{(j)}$ is the correction of order $\lambda^j$ to the time evolution operator. 
For a given initial detectors-field state $\roo=\rod \otimes \rof$, the final state of the detectors (after tracing out the field) is
\begin{align}
    \rd &= \tr_\phi \left(U \roo U^\dagger\right) = \rho_{\textsc{D}}^{(0)} + \rd^{(2)} + \mathcal{O}(\lambda^4),
\end{align}
where we have made use of the fact that the field is in the Minkowski vacuum, which is a Gaussian state, so all odd-point correlation functions vanish. Moreover, with $\Omega>0$, we assume the detectors start in a product state of their respective ground states, $\rod=\ketbra{0}_a \otimes \ketbra{0}_b$.

Unlike qubits, whose internal dynamics are fixed, qutrits admit richer internal dynamics (for instance, with or without superselection rules). In this work, we consider two internal dynamics for the qutrit detectors: SU(2) and Heisenberg-Weyl (HW), as prescribed in \cite{Lima:2023pyt}. The results presented in the main text focus on SU(2) dynamics, and the comparison with the HW case is included in \autoref{app:comparisonSU2HW}.

For the case of SU(2) detectors, the energy levels in Eq.~\eqref{eq:freeHamiltonian} and the detectors' observable coupling to the field amplitude in Eq.~\eqref{eq:interaction_h} are
\begin{align}
    \begin{cases}
        \Omega_{i,0}=0,\; \Omega_{i,1}=\Omega,\; \Omega_{i,2}=2\Omega \\
        O_i = \ketbra{0}{1}_i+ \ketbra{1}{2}_i + \text{h.c.}
    \end{cases}
\end{align}
In order to write the final state of the detectors, let us define
\begin{subequations}\label{eq:definitions_terms}
    \begin{align}
        \mathcal{L}_{ij} =&\, \lambda^2 \!\!\int\!\dd\mathsf{x} \,\dd\mathsf{x}' \Lambda_i(\mathsf{x}) \Lambda_j(\mathsf{x}') e^{-\iu  \Omega ( t - t')} W(\mathsf{x},\mathsf{x}'),\\
        \mathcal{Q}_i =& - \!\frac{\lambda^2}{2} \!\! \int\!\dd\mathsf{x} \, \dd\mathsf{x}'  \Lambda_i(\mathsf{x}) \Lambda_i(\mathsf{x}') e^{\iu \Omega (  t +t' )} G_F(\sx,\sx'),\\
        \mathcal{M}_{ab} =& - \lambda^2\!\! \int\!\dd\mathsf{x} \,\dd\mathsf{x}'  \Lambda_a(\mathsf{x}) \Lambda_b(\mathsf{x}') e^{\iu \Omega (  t +t' )} G_F(\sx,\sx'),
\end{align}
\end{subequations}
where $W(\sx,\sx')$ is the Wightman function,
\begin{align}
    W(\sx,\sx')&= \bra{0}\phi(\sx)\phi(\sx')\ket{0} \nonumber\\
    &=\int\frac{\dd^3\bk}{ 2 (2\pi)^3 |\bk|}\, e^{-\iu |\bk| (t-t') + \iu \bm{k}\cdot(\bx-\bx')}, 
\end{align}
and $G_{F}(\sx,\sx')= \Theta(t-t')W(\mathsf{x},\mathsf{x}') + \Theta(t'-t)W(\mathsf{x}',\mathsf{x})$ is the Feynman propagator of the field. Furthermore, given that we consider identical detectors, we write $\mathcal{L}\coloneqq\mathcal{L}_{aa}=\mathcal{L}_{bb}$ and $\mathcal{Q}\coloneqq \mathcal{Q}_a=\mathcal{Q}_b$.

Up to order $\lambda^2$ and in the ordered basis $\{ \ket{0}_a,\ket{1}_a,\ket{2}_a\}\otimes\{ \ket{0}_b,\ket{1}_b,\ket{2}_b\}$, the final state of the detectors is
\begin{equation}
    \rd =
    \begin{pmatrix}
        1-2\mathcal{L} & 0 & \mathcal{Q}^* & 0 & \mathcal{M}_{ab}^* & 0 & \mathcal{Q}^* & 0 & 0 \\
        0 & \mathcal{L} & 0 & \mathcal{L}_{ab} & 0 & 0 & 0 & 0 & 0 \\
        \mathcal{Q} & 0 & 0 & 0 & 0 & 0 & 0 & 0 & 0 \\
        0 & \mathcal{L}_{ab} & 0 & \mathcal{L} & 0 & 0 & 0 & 0 & 0 \\
        \mathcal{M}_{ab} & 0 & 0 & 0 & 0 & 0 & 0 & 0 & 0 \\
        0 & 0 & 0 & 0 & 0 & 0 & 0 & 0 & 0 \\
        \mathcal{Q} & 0 & 0 & 0 & 0 & 0 & 0 & 0 & 0 \\
        0 & 0 & 0 & 0 & 0 & 0 & 0 & 0 & 0 \\
        0 & 0 & 0 & 0 & 0 & 0 & 0 & 0 & 0
    \end{pmatrix},
\end{equation}
where we have used the fact that $\mathcal{L}_{ab}$ is real.

We consider the center of mass of detector $a$ to coincide with the origin, $(t, \bm x_a) = (0, \bm 0)$, and the distance $d$ between the two detectors to be fixed, so that the center of mass of detector $b$ is $(t, \bx_b) =(0, \bm d)$. 
Under these assumptions, the spacetime smearing can be written as a product of a switching function and a spatial smearing function, $\Lambda_i(\sx) = 
\chi_i (t) F_i(\bm x)$. Specifically, we use
\begin{subequations}\label{eq:switching_and_smearing}
    \begin{align}
        \chi_i(t)&= \exp\left( -\frac{t^2}{T^2} \right),\\
        F_i(\bm{x}) &= \frac{3}{4\pi R^3} \Theta \left( R - |\bm{x}-\bm{x}_i| \right),
    \end{align}
\end{subequations}
where $\Theta$ is the Heaviside step function; $T$ controls the effective duration of the switching, and $R$ is the radius of the spherical spatial smearing.

Although this switching function is not compactly supported, so that the two detectors are never strictly in the spacelike regime, we can define an intuitive criterion for an effective spacelike separation. The criterion relies on the observation that, for a Gaussian switching function centered at $t=0$, most of its weight is concentrated in the interval $[-\frac{N T}{2\sqrt{2}}, \frac{N T}{2\sqrt{2}} ]$, where $T/\sqrt{2}$ is the standard deviation of the Gaussian switching and $N\geq2$.

While we still integrate over $\mathbb{R}$, this observation allows us to treat the interaction of each detector as having an effective finite duration of $N\;T/\sqrt{2}$, resulting in the following criterion:
\begin{align}\label{eq:spacelikecriterion}
    d \geq 2R + N \frac{T}{\sqrt{2}}.
\end{align}
Notice that for near-pointlike detectors, this criterion reduces to the one more frequently seen in the literature, $d \geq N\;T/\sqrt{2}$. However, since we also explore large detector regimes, where the radius is comparable to the interaction time, $R$ must be taken into account. In our analysis, we will consider different values of $N$ for increasingly greater spacelike separations. In particular, the least stringent case used corresponds to $N=5$, for which over 98\% of the switching's weight lies within $\pm5T/\left(2\sqrt{2}\right)$ from its peak. In the plots, we will also set $N=7$, as frequently used in the literature~\cite{Pozas-Kerstjens:2015gta}, as well as $N=10$ and $N=14$.

The expressions for the terms in the final density matrix, with the spatial profile and switching chosen in Eq.~\eqref{eq:switching_and_smearing}, can be found in \autoref{app:integrals}.

\section{Contextuality}\label{sec:contfraction}
Contextuality captures the failure of a hidden-variable model to assign predetermined outcomes to all measurements in a system such that the outcomes are independent of the measurement context. Here, a measurement context is a set of measurements that are jointly implementable (or operationally compatible), such as a set of commuting projective measurements.
Consider the scenario where one attempts to give to the measurement of each observable a pre-written “answer” that does not depend on what else is measured alongside it. Whenever a set of observables can be jointly measured, those answers must fit together and respect the same algebraic relations and coarse-grainings that the measurements obey. 
\emph{Noncontextuality} would require the value attributed to each observable to be the same no matter which compatible measurements accompany it. However, quantum predictions show that this independence generally cannot hold. We refer the reader to Ref.~\cite{mermin_hidden_1993} for a pedagogical review on how noncontextual hidden-variables models fail to describe certain quantum systems.

One may try to emulate the probabilistic outcomes of quantum measurements by considering a probabilistic combination of pre-written answers. While this strategy may work for a subset of contexts, one can find systems with measurement sets where no such strategy can consistently reproduce the quantum measurements statistics on \textit{all} the contexts. These notions of context dependence are formalized in Ref.~\cite{abramsky_sheaf-theoretic_2011} in terms of sets of measurements and probability distributions, with appropriate marginalizations over them.

Formally, a measurement scenario is specified by a set of observables $\mathcal{A}$, a family of contexts $\mathcal{C}$, and the set of possible measurement outcomes $\mathcal{O}$. $\mathcal{C}$ is a subset of the power set of $\mathcal{A}$,   $\mathcal{C}\subseteq2^{\mathcal{A}}$, and  consists of subsets of mutually compatible observables. For a context $C$ in $\mathcal{C}$, $p_C$ is a vector that describes the statistics of outcomes for the joint measurement of all observables in $C$. Then, an empirical model $\mathcal{E}$ on a measurement scenario is a collection of probability distributions $\{p_C\}_{C \in \mathcal{C}}$. An empirical model is valid only when it satisfies the compatibility criteria, that is, the probability distributions over different contexts must be consistent when marginalized to the overlaps between the contexts \cite{abramsky_sheaf-theoretic_2011}. A model is said to be noncontextual if the statistics for each jointly measurable context arise as marginals of a single joint distribution~\cite{abramsky_sheaf-theoretic_2011, acin_combinatorial_2015, de_silva_graph-theoretic_2017}. If no such global distribution is possible, the empirical model is said to be contextual. 

Now, we introduce \emph{contextual fraction}, which quantifies the extent of contextuality in a given empirical model~\cite{abramsky_sheaf-theoretic_2011,abramsky_contextual_2017}.
Any empirical model $\mathcal{E} =  \{p_C\}_{C\in\mathcal{C}}$ can be expressed as a convex sum of two empirical models
\begin{equation}\label{eq:convex_decomp}
    \mathcal{E} = p \,\mathcal{E}_{\text{NC}} + (1-p)\,
    \mathcal{E}^{\prime},
\end{equation}
where $\mathcal{E}_{\text{NC}}$ is a noncontextual model and $\mathcal{E}^{\prime}$ an arbitrary (possibly contextual) model. Note that any empirical model admits such a convex decomposition, and the decomposition need not be unique. For a convex decomposition of this form, when the weight $p$ of the noncontextual part is maximal, the weight $(1-p)$ of the remaining model is the contextual fraction $CF(\mathcal{E})$ of the empirical model. This provides a robust measure of contextuality as it quantifies the fraction of the measurement statistics that cannot be explained in terms of noncontextual models~\cite{abramsky_sheaf-theoretic_2011}.

Operationally, computing the contextual fraction reduces to a linear optimization problem~\cite{abramsky_contextual_2017}. For an empirical model $\mathcal{E}$, let $v_{\mathcal{E}}$ be the vectorized empirical model obtained by stacking the probability vectors of all the contexts.  
If $\mathcal{E}$ is noncontextual, then $v_\mathcal{E}$ can be expressed as a convex combination of $\{v_d\}_{d \in \mathcal{D}}$, where $\mathcal{D}$ is the set of all deterministic global assignments, and $v_d$ are the vectorized empirical models generated by these assignments. Here, a \emph{deterministic global assignment} is a map that picks a single outcome in $\mathcal{O}$ for every measurement of an observable in $\mathcal{A}$. The condition for noncontextual $\mathcal{E}$ can be stated as a linear constraint: $v_{\mathcal{E}} = \mathsf{M}. b$, where the columns of the matrix $\mathsf{M}$ are the empirical models $v_d$, and $b$ is a vector of probabilities specifying the convex decomposition. However, for a general empirical model, we can rewrite the decomposition in Eq.~\eqref{eq:convex_decomp} as
\begin{equation}\label{eq: vectorized_decomposition}
    v_{\mathcal{E}} =  \mathsf{M}.b^{\prime} + (1-\sum_i b\superprime_i)\,v_{\mathcal{E}^{\prime}},
\end{equation}
where $b^{\prime}$ is a \emph{subprobability distribution} into which we have absorbed the weight of the noncontextual contribution, $b\superprime_i$ are the entries of $b\superprime$, and $v_{\mathcal{E}'}$ is the vectorized empirical model for $\mathcal{E}'$. Using Eq.~\eqref{eq: vectorized_decomposition}, the contextual fraction of the empirical model
\begin{equation}
CF(\mathcal{E}) = 1 - \sum_{i} b'_i
\end{equation}
can be computed by finding the vector $b^{\prime}$ such that the sum of its entries is maximized.

We evaluate the contextual fraction for the measurement statistics of the final detector state $\rd$ with respect to the two-qutrit Heisenberg-Weyl measurement scenario. The single-qutrit Heisenberg-Weyl group is generated by the clock and shift operators, $Z$ and $X$~\cite{gottesman_stabilizer_1997}, which are defined as
\begin{align}\label{eq:clock_and_shift_operators}
    Z = \sum_{j=0}^{2} \omega^j \ketbra{j}{j}\,,\quad
    X &= \sum_{j=0}^{2}\ketbra{j+1\!\!\!\!\mod 3}{j},  
\end{align}
where $\omega = e^\frac{2\pi \ii}{3}$.
The elements of the single-qutrit Heisenberg-Weyl group can be written as
\begin{equation} \label{eq: HW_elements_1q}
   W(\bm v)= W([p, q])= \omega^{2^{-1} p q} X^p Z^q, 
\end{equation}
where $\bm v = [p,q] \in \Z_3^2$ denotes a phase point vector and $2^{-1}$ is the multiplicative inverse of $2$ in $\Z_3$. For two qutrits, the group elements are identified by elements in $\Z_3^4$, and take the tensor-product form
\begin{equation}\label{eq: HW_elements_2q}
 W[p_1,q_1,p_2,q_2] = W[p_1,q_1] \otimes W[p_2,q_2].  
\end{equation}

Note that the Heisenberg-Weyl operators are non-Hermitian. However, here, “observables” for the Heisenberg–Weyl group are understood in the standard spectral sense: each unitary \(W(\mathbf v)\) is measured via its projection-valued measure, i.e., the eigenprojectors $\{\Pi^{(\mathbf k)}_v\}_{k\in\mathbb Z_3}$ for eigenvalues $\omega^k$. Outcomes are therefore labels \(k\in\mathbb Z_3\), and all probabilities are computed with these projectors. 
The compatibility of the observables is captured by their commutation relations: two Heisenberg-Weyl operators are co-measurable if they commute. To characterize when these operators commute, it is convenient to equip the space of phase point vectors $\Z_3^4$ with a symplectic form that encodes their commutation relations \cite{hostens_stabilizer_2005}. For phase point vectors $\bm v = [p_1, q_1, p_2, q_2]$ and $\bm v\superprime = [p_1\superprime , q_1\superprime, p_2 \superprime, q_2 \superprime]$, the symplectic form is defined as 
\begin{equation}\label{eq: sym_product}
    [\bm v,\bm {v\superprime}] = p_1q_1\superprime - q_1p_1\superprime + p_2q_2\superprime - q_2p_2\superprime \in \Z_3, 
\end{equation}
and the commutation relations between these observables can be stated in the compact form
\begin{equation}\label{eq: HW_compose}
    W(\bm v)W(\bm {v\superprime}) = \omega^{2[\bm v,\bm{v\superprime}]}W(\bm v\superprime) W(\bm{v}). 
\end{equation}
Hence, two Heisenberg-Weyl operators on two qutrits, $W(\bm v)$ and $W(\bm{v\superprime})$, commute if their phase point vectors are orthogonal with respect to the symplectic form in Eq.~\eqref{eq: sym_product}, that is, $[\bm v,\bm{v\superprime}] = 0$. 
Therefore, the contexts of the Heisenberg-Weyl measurement scenario correspond to the subspaces of $\Z_3^4$ where the symplectic form vanishes, namely, the maximal isotropic subspaces.
Each isotropic subspace has nine elements. Equivalently, each context has nine operators: eight non-trivial and the identity. In total, there exist 40 contexts.

With the contexts identified this way, we compute the empirical probabilities by defining projectors onto the common eigenbases of the contexts. Specifically, for every set of mutually commuting Heisenberg–Weyl operators, we construct projectors corresponding to joint measurements of the context as 
\begin{equation}\label{eq: context_projecto}
    \Pi_C^{\bm{r}} = \frac{1}{9}\sum_{W \in C} \omega^{-r(W)}W.
\end{equation}
Here, the vector $\bm{r} \in \mathbb{Z}_3^9$ labels the outcome of the joint measurement where the operator $W \in C$ outputs $\omega^{r(W)}$, with $r(W)$ being the component of $\bm{r}$ labelled by $W$. Using these projectors, we can calculate all the outcome probabilities as
\begin{equation}\label{eq: born_rule}
    p_C(\bm{r}) = \Tr{\Pi_C^{\bm r} \rd},
\end{equation}
which yield the joint outcome distributions of our empirical model.

The constraint matrix $\mathsf{M}$ is generated by enumerating all the possible deterministic global assignments $\Lambda: \Z_4^3 \rightarrow \Z_3$
for the two-qutrit Heisenberg-Weyl group, where $\omega^{\Lambda(\bm v)}$ is the measurement outcome assigned to the operators $W(\bm v)$. Noncontextuality requires that these assignments must respect all algebraic relations among jointly measurable observables. Hence, for $[\bm v, \bm v\superprime] = 0$, the value assignments must follow
 \begin{equation}\label{eq: assignment_constratints}
     \Lambda(\bm v) + \Lambda (\bm v\superprime) = \Lambda(\bm v + \bm v\superprime).
 \end{equation}

In Ref.~\cite{delfosse_equivalence_2017}, the authors show that for the Heisenberg-Weyl group of two qutrits, the above constraint forces the deterministic value assignments to be linear in $\bm v \in \Z_3^4$, i.e., any such global assignment $\Lambda$ must be of the form $\Lambda(\bm v) =  \bm \lambda \cdot \bm v$ for some $\lambda \in \Z_3^4$.
This fact keeps the optimization tractable: our constraint matrix $\mathsf{M}$ has dimensions $360 \times 81$, corresponding to 9 outcomes for each of 40 contexts, and 81 global assignments.  
The optimum subprobability distribution that captures the noncontextual contribution to this empirical data is numerically obtained using the \texttt{linprog} optimizer in Python's \texttt{scipy.optimize} module \cite{virtanen_scipy_2020}. The contextual fraction of the state, $CF(\rd)$, can then be computed as $1-\sum_i b\superprime_i$, where $b\superprime$ is the optimal solution of the linear optimization\footnote{A repository of the Python code used for the numerical simulations is available \href{https://github.com/Sanchit-Srivastava/contextuality_from_the_vacuum}{here}.}.

Finally equipped with the necessary definitions, we comment on why the protocol uses qutrits rather than the more commonplace qubit UDW detectors, as promised at the beginning of \autoref{sec:setup}. The choice of qutrit is motivated by the fact that for two-qubit systems, the Heisenberg-Weyl measurement scenario yields state-independent contextuality: \textit{all} two-qubit states are contextual with respect to Heisenberg-Weyl measurements \cite{mermin_hidden_1993, raussendorf_contextuality_2017}. As a result, a measure of contextuality cannot distinguish between two-qubit states, nor can it capture the emergence of contextuality under interaction with the field. By contrast, three-level systems exhibit state-dependent contextuality in the two-qutrit Heisenberg-Weyl measurement scenario, making qutrit detectors the minimal choice for our analysis. We note that this measurement scenario also rules out an analysis with only one qutrit detector interacting with the field, as a single qutrit state is always noncontextual with respect to the single qutrit Heisenberg-Weyl group \cite{Howard:2014zwm}.

\section{Wigner negativity}\label{sec:Wignernegativity}

For a system of multiple qudits, stabilizer states are the pure states uniquely defined (up to phase) as the $+1$ simultaneous eigenstate of a maximal Abelian subgroup of the generalized Pauli group \cite{gottesman_stabilizer_1997}. The operations that map the set of stabilizer states to itself form the Clifford group. By the Gottesman-Knill theorem, any circuit composed of Clifford operations acting on a stabilizer input state is efficiently classically simulable \cite{aaronson_improved_2004}. To achieve universal quantum computation, this classically tractable framework must be supplemented with a non-Clifford resource, most commonly a non-stabilizer input state. Such states, which are essential for quantum speedup, are known as \textit{magic states} \cite{bravyi_universal_2005}.

Magic is therefore manifested in quantum states that lie outside the convex hull of stabilizer states (i.e., the \textit{stabilizer polytope}). For systems of odd prime dimension (such as the qutrits, $d=3$, considered here), the discrete Wigner function provides a natural framework to quantify this resource. States with non-negative Wigner functions form a convex set $\mathcal{W}_+$ that includes the stabilizer polytope. Crucially, Wigner negativity (i.e., the presence of negative values in this quasi-probability distribution) is a necessary condition for a state to be magic. It quantifies the deviation from this set of classically simulable states and serves as a verifiable magic monotone \cite{veitch_negative_2012}.

In multiple studies~\cite{delfosse_equivalence_2017,spekkens_negativity_2008,Howard:2014zwm}, contextuality and negativity of the Wigner function are shown to be equivalent notions of non-classicality. Therefore, in addition to evaluating the contextual fraction of the two-qutrit detector final state, $\rd$, we compute the discrete-Wigner negativity of the reduced state of one detector, $\rho_A$, as a consistency check.
While the ordinary Wigner function lives on a continuous phase space, here the phase space is discrete and forms a $3\times3$ grid for a single qutrit.
We index the nine phase-space points by vectors $\bm r=(x,y,x{+}y,x{+}2y)$ with $x,y\in\mathbb Z_3$. Each point has an associated phase-point operator $A^{\bm r}$ defined by~\cite{gross_hudsons_2006}
\begin{equation}\label{eq: phase_operators}
    A^{\bm r} = -\openone_3 + \sum_{i=1}^{4} \Pi_i^{r_i},
\end{equation}
where $\Pi_i^{r_i}$ projects onto the eigenvector with eigenvalue $\omega^{r_i}$ of the $i$-th operator in $\{W[0,1],\,W[1,0],\,W[1,1],\,W[1,2]\}$.
For any density operator $\rho$, the discrete Wigner value at $\bm r$ is
\begin{equation}
    \mathcal W_{\bm r}(\rho)=\tfrac{1}{3}\Tr\!\big(\rho\,A^{\bm r}\big).
\end{equation}

The set
\[
\mathcal W_+ \coloneqq \big\{\rho:\ \mathcal W_{\bm r}(\rho)\ge 0\ \ \forall\,\bm r\big\}
\]
is a convex polytope in state space (finitely many linear facet inequalities $\Tr(\rho A^{\bm r})\ge 0$). For odd $d$, $\mathcal W_+$ strictly contains the stabilizer polytope (the convex hull of stabilizer states); this region is the \textit{Wigner polytope}. The Discrete Hudson’s theorem implies that pure states with non-negative Wigner function are exactly the stabilizers; however, Gross showed that there exist non-negative \textit{mixed} states that are not stabilizer states~\cite{gross_hudsons_2006}.

Following Veitch \emph{et al.} \cite{veitch_negative_2012}, we quantify magic by the sum-negativity
\begin{equation}\label{eq: wigner_negativity}
    N(\rho) = \sum_{\bm r:\,\mathcal W_{\bm r}(\rho)<0}\big|\mathcal W_{\bm r}(\rho)\big|,
\end{equation}
a magic monotone for odd $d$; non-negativity $(N(\rho)=0)$ is a known obstruction to magic-state distillation and enables efficient classical simulation of stabilizer circuits supplied with arbitrary copies of $\rho$.

For the final detector state $\rho_{\mathcal D}$, the reduced state of the first detector is
\begin{align}
    \rho_A &= \begin{pmatrix}
        1-\mathcal{L} & 0 & \mathcal{Q}^*\\
        0 & \mathcal{L} & 0\\
        \mathcal{Q} & 0 & 0
    \end{pmatrix}.
\end{align}
Evaluating the nine Wigner values $\mathcal W_{\bm r}(\rho_A)$ shows (by elementary positivity constraints on $\rho_A$) that only three can be negative; up to the common factor $\tfrac{1}{3}$ they are
\begin{subequations}\label{eq:inequalities}
\begin{align}
   \mathcal{L} + 2\,\Re[\mathcal{Q}], \label{eq:lessviolatedinequality}\\
   \mathcal{L} - \Re[\mathcal{Q}] + \sqrt{3}\,\Im[\mathcal{Q}],\\
   \mathcal{L} - \Re[\mathcal{Q}] - \sqrt{3}\,\Im[\mathcal{Q}]. \label{eq:moreviolatedinequality}
\end{align}
\end{subequations}
In our parameter regime, only \eqref{eq:lessviolatedinequality} and \eqref{eq:moreviolatedinequality} become negative, yielding $N(\rho_A)>0$. We observe that these instances coincide with a nonzero contextual fraction for the corresponding two-qutrit state $\rho_{\mathcal D}$. This is consistent with the equivalence (for odd local dimension) between state-dependent contextuality for Pauli/stabilizer measurements and Wigner-function negativity~\cite{delfosse_equivalence_2017}.

\section{Results}\label{sec:results}

\begin{figure*}[t]
    \centering

    \subfloat[]{%
        \includegraphics[width=0.48\textwidth]{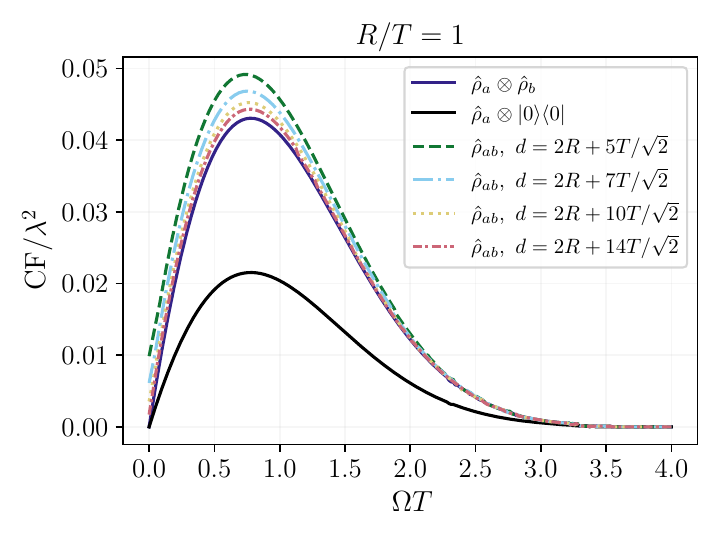}%
        \label{fig:contextual_fraction_vs_gap_large_detectors}%
    }\hfill
    \subfloat[]{%
        \includegraphics[width=0.48\textwidth]{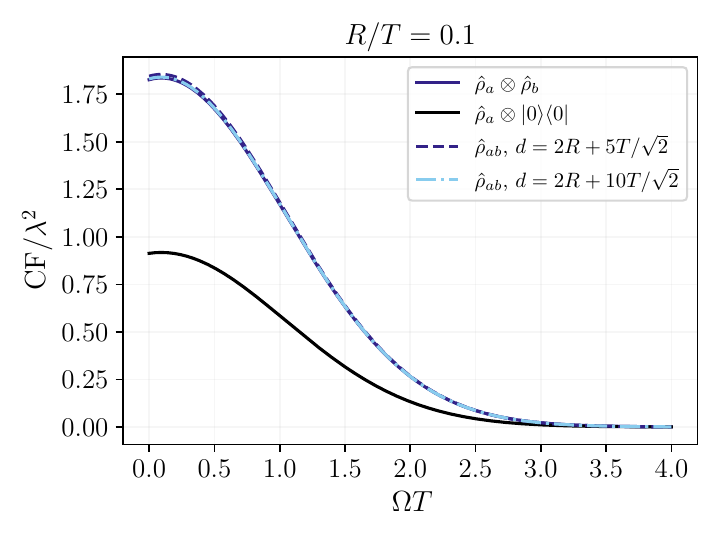}%
        \label{fig:contextual_fraction_vs_gap_small_detectors}%
    }

    \subfloat[]{%
        \includegraphics[width=0.48\textwidth]{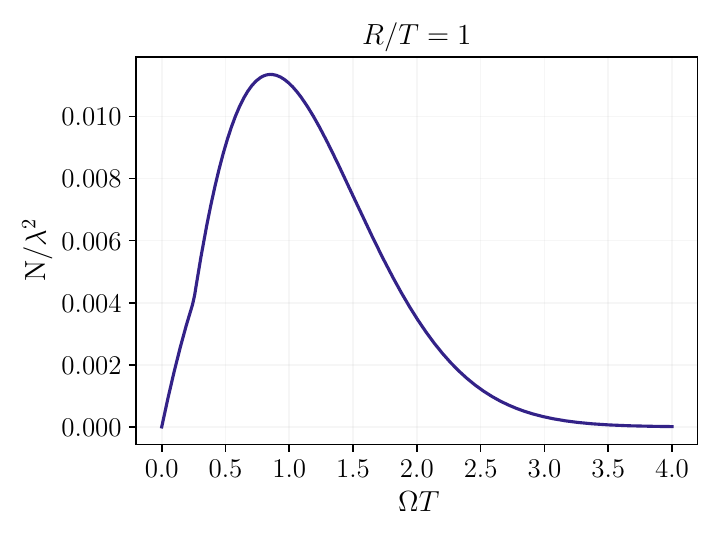}%
        \label{fig:magic_vs_gap_big_detectors}%
    }\hfill
    \subfloat[]{%
        \includegraphics[width=0.48\textwidth]{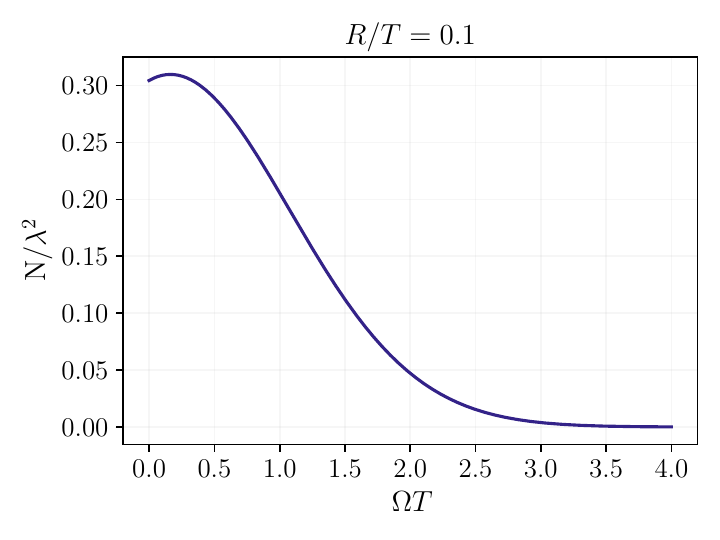}%
        \label{fig:magic_vs_gap_small_detectors}%
    }

    \caption{Contextual fraction (in (a)–(b)) and Wigner negativity (in (c)–(d)) normalized by the coupling constant squared as a function of the product of the detectors' gap and their interaction duration. As compared to the interaction time scale, the detector is ``large'' in (a) and (c) and ``small'' in (b) and (d).}
    \label{fig:mainplots}
\end{figure*}

Previous work \cite{Nystrom:2024oeq} has shown that a single Unruh–DeWitt detector interacting with the vacuum of a quantum field can acquire magic, quantified by mana. Since the results of Refs.~\cite{Howard:2014zwm} and \cite{delfosse_equivalence_2017} establish that nonzero magic necessarily implies nonzero contextuality, one can expect that contextuality harvesting is already guaranteed in such scenarios. However, these equivalence results are qualitative: they certify the presence of contextuality but do not determine its magnitude or how it depends on the physical parameters of the interaction. The results presented in this section include both the magnitude and parameter dependence of the contextuality harvested by two qutrits; furthermore, they help address a natural question: what advantage does our protocol have over harvesting magic with a single qutrit?  

To answer this question, we compare the total contextual fraction harvested in two settings. Suppose Alice and Bob each begin with a qutrit in the ground state, and consider two possible scenarios: (i) both detectors couple to the quantum field and carry out our contextuality-harvesting protocol; (ii) only Alice couples her detector to the field, following the magic-harvesting protocol, while Bob’s detector remains uncoupled. For scenario (i), we run our two-detector protocol for various values of separation $d$ (measured in units of $T$), restricting to the cases which satisfy our criterion for spacelike separation in Eq. \eqref{eq:spacelikecriterion}. Since the final state obtained in the magic-harvesting protocol is indeed the reduced state of a single detector in our setup, we denote the final detector state in scenario (ii) as $\rd = \ra \otimes \ketbra{0}$. For comparison, we also include the limiting case which corresponds to the asymptotic regime at large separations, where the joint detector state approaches the product of its single-detector reductions. We label this state as $\rd = \ra \otimes \rb$.

In Figs. \ref{fig:contextual_fraction_vs_gap_large_detectors} and \ref{fig:contextual_fraction_vs_gap_small_detectors}, we show the plots of contextual fraction $CF(\rd)$ normalized by $\lambda^2$ as a function of the adimensional parameter $\Omega T$. Two detector sizes are considered: (a) a large detector with radius $R=1$ (in units of $T$), and (b) a smaller detector, closer to the pointlike regime. In both cases, the contextual fraction displays an optimal energy gap at which it is maximum, and this optimal value shifts depending on the detector size. Notably, for $R=0.1T$ the optimal gap is smaller than that for $R=T$. In fact, given that gapless detectors can not harvest entanglement~\cite{pozas-kerstjens_degenerate_2017}, a surprising conclusion in our study is that for the $R=0.1T$ setup the optimal gap can be very close to zero, as seen in Fig. \ref{fig:contextual_fraction_vs_gap_small_detectors}. 
Another observation is that the maximum harvested contextual fraction increases as the detector size is decreased relative to the interaction time, i.e., as we approach the pointlike limit. This trend mirrors observations in entanglement harvesting \cite{Pozas-Kerstjens:2015gta}. 

Across all configurations, we restrict to setups within the spacelike separation regime, according to the criterion in Eq. \eqref{eq:spacelikecriterion}. As the detectors' separation increases, the contextual fraction decreases and asymptotically approaches the value for the tensor product of the single-detector states. While the difference between the curves and this asymptote is more pronounced in the large-detector case (a), the same phenomenon persists for the small detector case (b), though less apparent in the shown plots. Finally, we note that the contextual fraction harvested in scenario (i) with two detectors is consistently larger than that obtained in the single-detector magic-harvesting protocol in scenario (ii), even in the limit of asymptotically large separations between Alice and Bob. 

In Figs.~\ref{fig:magic_vs_gap_big_detectors} and \ref{fig:magic_vs_gap_small_detectors} we plot the Wigner negativity of the reduced detector state normalized by $\lambda^2$, corresponding to the contextual fraction plots for the same detector size in Figs. \ref{fig:contextual_fraction_vs_gap_large_detectors} and \ref{fig:contextual_fraction_vs_gap_small_detectors}  respectively. As expected from the theory, the shapes of the curves for the harvested contextual fraction and Wigner negativity are similar. 
Some comments regarding the behaviour of the $CF$ and $N$ in Fig.~\ref{fig:mainplots} are in order. First, as noted before, $\Omega T =0$ corresponds to gapless detectors ($\Omega=0$) and not to a vanishing interaction time ($T=0$) because $R/T$ is a fixed finite number. Consequently, the contextual fraction and Wigner negativity do not have to vanish when $\Omega T=0$; and indeed they do not in Figs.~\ref{fig:mainplots}a, b and c. In \autoref{app:more_plots}, we provide complementary results obtained by fixing $R\Omega$ (instead of $R/T$), in which case $\Omega T=0$ corresponds to a vanishing interaction time. As a consequence, both $CF$ and $N$ vanish, as expected for $T=0$, in agreement with the results reported in Ref.~\cite{Nystrom:2024oeq}.

Second, we observe that the $CF$ and $N$ drop to zero for large $\Omega T$ in Fig.~\ref{fig:mainplots}; equivalently, we see in Eqs.~\eqref{eq:Lcal_dimensionless}-\eqref{eq:final_integrals} that all the $\mathcal{O}(\lambda^2)$ contributions to $\hat{\rho}_D$ are exponentially suppressed in $\Omega T$. This suppression for large $\Omega T$ boils down to the uncertainty principle as follows. The expressions in Eq.~\eqref{eq:definitions_terms}, after some manipulations, involve products of (ordinary or one-sided) Fourier transforms of the Gaussian switching functions with respect to $(t-t')$ or $(t+t')$. As per the uncertainty principle, if the Gaussian functions in time-space have a variance proportional to $T$, then their Fourier transforms in frequency-space have a variance proportional to $1/T$. Therefore, a longer interaction duration results in a smaller contributing frequency window, suppressing the $\mathcal{O}(\lambda^2)$ terms. For instance, it is well known that an inertial detector, initially in its ground state, and with a Gaussian switching function, has a non-zero transition probability $\mathcal{L}$ that vanishes in the $T\to\infty$ limit~\cite{birrell_quantum_1982, sriramkumar_finite-time_1996,satz_then_2007}.

Third, the non-zero $CF$ and $N$ for small $\Omega T$ can also be traced back to the uncertainty principle: short interaction durations result in broad contributing frequency windows, making the $\mathcal{O}(\lambda^2)$ terms non-vanishing.
Therefore, competitions between the $\mathcal{O}(\lambda^2)$ terms, like those given in Eq.~\eqref{eq:inequalities}, along with the interplay of the details of the Wightman, the switching, and smearing functions, determine the non-monotonic behaviour of the $CF$ and $N$ for small $\Omega T$. 

An important check on our results concerns their dependence on the coupling strength $\lambda$. From perturbative considerations, one expects the harvested contextuality to scale as $\lambda^2$, provided we remain in the regime where $O(\lambda^4)$ corrections are negligible compared to leading $O(\lambda^2)$ terms. It is not immediately clear, however, whether this scaling should persist through the linear optimization used to compute the contextual fraction. Our simulations show that for $10^{-4} < \lambda <10^{-2}$, the ratio $CF(\rd)/\lambda^2$ is indeed constant with varying $\lambda$. For larger couplings $\lambda > 10^{-2}$, we interpret the deviations we see in our simulations as signatures of higher-order corrections becoming relevant. On the other hand, for very small couplings $\lambda < 10^{-4}$, deviations from the constant behaviour can be attributed to instability of the numerical simulations due to floating point errors. We posit that with an exact solver and sufficiently high-precision integrations, the expected constant behaviour would extend to this regime. Our plots are reported with $\lambda = 10^{-3}$, which lies safely within the range where the $\lambda^2$ scaling is reliable.

\section{Discussion}\label{sec:discussion}
In this article, we propose an extension of the canonical entanglement harvesting protocol to study the harvesting of another resource: contextuality. Concretely, we show that an initially noncontextual two-qutrit state acquires contextuality after locally interacting with the vacuum of a quantum field. The set of measurements with respect to which we show contextuality is the Heisenberg-Weyl group of two qutrits, and the measure of contextuality used is the contextual fraction.

A greater contextual fraction signifies a more pronounced departure from classical hidden variable models, which is directly linked to the state's resilience and potential for enhanced performance in tasks like quantum communication, randomness certification, etc \cite{Amaral2019, gupta_quantum_2023, schmid_contextual_2018, hu_self-testing_2023, um_experimental_2013, um_randomness_2020}. This suggests that tailoring the interactions with the field could be a viable method for engineering states not just to be magical, but to be optimally equipped for specific quantum information processing applications where a greater degree of non-classicality is a resource.

Possible extensions of our work are, for instance, investigations in flat spacetime dimensions different than 1+3, curved spacetimes, coupling the qutrits to other quantum fields, or considering higher-dimensional odd prime qudits. Moreover, we have not surveyed our parameter space for optimization, so an interesting proposal would be to understand what ranges for each detector parameter would increase the contextual fraction harvested, in the same vein as done in Ref.~\cite{Pozas-Kerstjens:2015gta} for entanglement harvesting. A comparison between the optimal detector parameters for harvesting entanglement and those for harvesting contextuality would be particularly interesting, since some of our findings mirror the behavior observed in harvested entanglement, while others deviate from it. For instance, we find that, keeping $R/T$ fixed (i.e., for finite size detectors), gapless detectors can harvest contextuality and magic, which is not the case when the resource is entanglement \cite{pozas-kerstjens_degenerate_2017}. On the other hand, larger detectors harvest less contextuality, which agrees with the case of entanglement \cite{Pozas-Kerstjens:2015gta}.

Another possible analysis would be a comparison between entanglement negativity and contextual fraction harvested in our protocol. Note, however, that although the two detectors generally become correlated---since the vacuum of a quantum field theory is a highly entangled state---and that we expect the two qutrits to have harvested entanglement, negativity fails to capture bound entanglement in two-qutrit systems, in contrast with the qubit–qubit and qubit–qutrit cases (where no bound entanglement is found). For (an upcoming) discussion of bound entanglement in two-qutrit UDW detectors, see Ref.~\cite{ParryPreciadoMann_inprep}. Despite this drawback related to negativity, our choice of qutrits rather than qubits is motivated by the fact that connections between contextuality in bipartite qudit systems and magic in their single-qudit reductions are only available when the local dimension is an odd prime.

The interaction Hamiltonian in the harvesting protocol is a separable operator; hence, any entanglement in the detectors must be field-mediated. This mediation can occur through two mechanisms: genuine extraction of pre-existing correlations from the field or communication through the field. The first one depends on the state of the field, while the second one is independent of it. A mixture of both mechanisms can occur for detectors in causal contact, but entanglement is guaranteed to be completely genuine (i.e., extracted from the field alone) for spacelike separated detectors~\cite{tjoaWhenEntanglementHarvesting2021}. Using the latter scenario, we can conclude that the field itself must be entangled at different points of spacetime. In the case of contextuality, we are unaware of an analogous argument that would imply that it is genuinely harvested from the field.
Thus, our results do not show that the field itself must be contextual either. Answering the questions of whether the contextual fraction depends on the state of the field and whether any local operations (independent of the field) increase contextuality in the detectors would help us establish the authenticity of contextuality harvesting, but we leave these as topics for future work.

Our results support an operational viewpoint in which the vacuum of a quantum field can be regarded as a ``free'' source of not only entanglement but also additional nonclassical resources.

\section*{Acknowledgements}
We thank Avantika Agarwal, Marina Maciel Ansanelli, Amolak Ratan Kalra, Robert B. Mann, Eduardo Martín-Martínez, Leo Parry, Everett Patterson, Erickson Tjoa, and the anonymous referees of the 29th Annual Quantum Information Processing Conference 2026 for insightful discussions and comments. We also thank the hospitality of Perimeter Institute and the Institute of Quantum Computing during the 2025 Quantum Information Winter Retreat (organized by Alex May), where part of this work was developed.
CL acknowledges the support from the Natural Sciences and Engineering Research Council of Canada through a Vanier Canada Graduate Scholarship [Funding Reference Number: CGV -- 192752].
MRPR gratefully acknowledges the support provided by the Mike and Ophelia Lazaridis Graduate Fellowship and her supervisor, Robert B. Mann.
SS acknowledges the support of his supervisors Shohini Ghose and Robert B. Mann, and the Natural Sciences and Engineering Research Council of Canada (NSERC). 
Research at Perimeter Institute is supported by the Government of Canada through the Department of Innovation, Science and Industry Canada and by the Province of Ontario through the Ministry of Colleges and Universities.

\bibliography{references}

\appendix
\begin{widetext}

\section{Comparison between SU(2) and Heisenberg-Weyl detectors}\label{app:comparisonSU2HW}

In this appendix, we describe the Heisenberg-Weyl internal dynamics, and compare them with the SU(2) case results detailed in the main text. For the HW qutrit, the energy levels in Eq.~\eqref{eq:freeHamiltonian} and the detector's observable coupling to the field amplitude are
\begin{align}
    &\begin{cases}
        \Omega_{i,0}=0,\; \Omega_{i,1}=\Omega_{i,2}=\Omega \\
        O_i = \ketbra{0}{1}_i + \ketbra{0}{2}_i + \ketbra{1}{2}_i + \text{h.c.}
    \end{cases}
\end{align} 
The nomenclature comes from the fact that we can build $h_{D,i}$ and $O_i$ a linear combination of the Heisenberg-Weyl generalized $X$ and $Z$ operators defined in Eq. \eqref{eq:clock_and_shift_operators},
and their products. Note that SU(2) has a superselection rule---no transitions occur between states $\ket{0}_i$ and $\ket{2}_i$--- which is absent in the HW qutrit. On the other hand, HW has two degenerate excited states when $\Omega>0$. For the final state of the detector in the HW case, we need one more definition, apart from the ones in Eqs. \eqref{eq:definitions_terms} of the main text:
\begin{align}\label{eq:V_i}
    \mathcal{V}_i =& - \lambda^2\!\! \int\!\dd\mathsf{x} \,\dd\mathsf{x}'  \Lambda_i(\mathsf{x}) \Lambda_i(\mathsf{x}') e^{\iu \Omega t'} \Theta(t-t')W(\sx,\sx').
\end{align}
The final state of the two detectors for the HW case can be written as
    \begin{align}
            \rd^{\text{HW}} \!\!=\!\!
    \begin{pmatrix}
        1 - 4\mathcal{L} & \mathcal{V}^* & \mathcal{V}^* & \mathcal{V}^* & \mathcal{M}_{ab}^* & \mathcal{M}_{ab}^* & \mathcal{V}^* & \mathcal{M}_{ab}^* & \mathcal{M}_{ab}^* \\
        \mathcal{V} & \mathcal{L} & \mathcal{L} & \mathcal{L}_{ab} & 0 & 0 & \mathcal{L}_{ab} & 0 & 0 \\
        \mathcal{V} & \mathcal{L} & \mathcal{L} & \mathcal{L}_{ab} & 0 & 0 & \mathcal{L}_{ab} & 0 & 0 \\
        \mathcal{V} & \mathcal{L}_{ab} & \mathcal{L}_{ab} & \mathcal{L} & 0 & 0 & \mathcal{L} & 0 & 0 \\
        \mathcal{M}_{ab} & 0 & 0 & 0 & 0 & 0 & 0 & 0 & 0 \\
        \mathcal{M}_{ab} & 0 & 0 & 0 & 0 & 0 & 0 & 0 & 0 \\
        \mathcal{V} & \mathcal{L}_{ab} & \mathcal{L}_{ab} & \mathcal{L} & 0 & 0 & \mathcal{L} & 0 & 0 \\
        \mathcal{M}_{ab} & 0 & 0 & 0 & 0 & 0 & 0 & 0 & 0 \\
        \mathcal{M}_{ab} & 0 & 0 & 0 & 0 & 0 & 0 & 0 & 0 \\
    \end{pmatrix},
    \end{align}
where we defined $\mathcal{V}\coloneqq\mathcal{V}_a=\mathcal{V}_b$ since our detectors are identical. Finally, the reduced state of a single detector is given by
\begin{align}
    \ra^{\text{HW}} &= \begin{pmatrix}
        1-2\mathcal{L} & \mathcal{V}^* & \mathcal{V}^*\\
        \mathcal{V} & \mathcal{L} & \mathcal{L}\\
        \mathcal{V} & \mathcal{L} & \mathcal{L}
    \end{pmatrix}.
\end{align}
Replacing $\mathcal{Q}$ with $\mathcal{V}$ in Eq. \eqref{eq:inequalities}, we obtain all the inequalities that can potentially be violated for the HW case. Similarly to what was observed for SU(2), our results show that for the reduced single-detector state, the only inequalities that can be violated for any range of parameters are Eqs.~\eqref{eq:lessviolatedinequality} and \eqref{eq:moreviolatedinequality}.

\begin{figure*}
    \centering
    \subfloat[]{
    \includegraphics[width=0.48\linewidth]{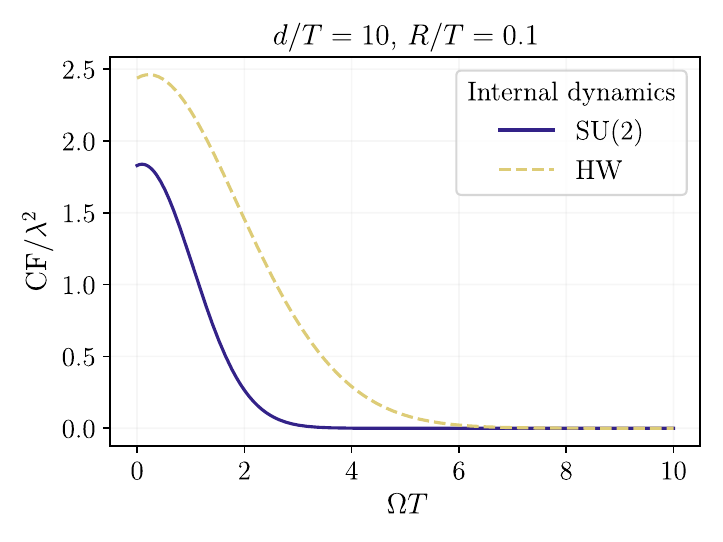}
    }\hfill
    \subfloat[]{
    \includegraphics[width=0.48\linewidth]{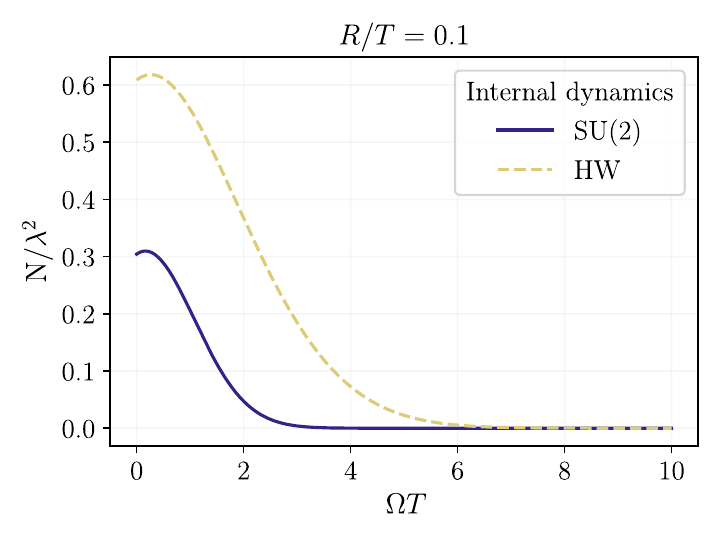}
    }
    \caption{Contextual fraction (a) and Wigner negativity (b) (normalized by the squared coupling constant) as a function of $\Omega T$, both for the SU(2) and for the Heisenberg-Weyl qutrits. The plot is for small detectors in the spacelike regime; note, however, that the Wigner negativity is independent of the distance between the two detectors.}
    \label{fig:comparisonSU2Hw}
\end{figure*}

To illustrate the effect of the qutrit's internal dynamics in harvesting contextuality and magic, we show in \autoref{fig:comparisonSU2Hw} the normalized contextual fraction (for the two-qutrit final state) and the normalized Wigner negativity (for the reduced single-qutrit state) as a function of $\Omega$, keeping $T=1$ constant. As discussed in the main text, the Wigner negativity and the contextual fraction follow the same trend for either internal dynamics. However, there are quantitative differences between the two prescribed internal dynamics. In fact, for most setups, the HW qutrit seems to be able to harvest more contextual fraction and Wigner negativity than the SU(2) particle detector model. This is likely related to two facts: first, the HW qutrit model has no superselection rule, allowing for direct transitions between $\ket{0}$ and $\ket{2}$; second, the HW model has a degenerate excited state. Combined, those two conditions imply that the states $\ket{1}$ and $\ket{2}$ have the same probability of becoming populated, which does not occur for SU(2).

\section{Integrals relevant to the final state of the detectors}\label{app:integrals}
Herein, we report the expressions for the integrals in Eqs.~\eqref{eq:definitions_terms} and \eqref{eq:V_i}, using the switching and smearing functions in Eq.~\eqref{eq:switching_and_smearing},
\begin{align}
    \mathcal{L} =&\; \alpha \!\int_{0}^{\infty}\! \frac{\dd k}{ k} \;j_1^2 (k \tilde{R})\;  e^{-\frac{1}{2} (k + \omega)^2}\!,\label{eq:Lcal_dimensionless}\\
    \mathcal{L}_{ab} =&\; \alpha \!
    \int_{0}^{\infty}\! \frac{\dd k}{ k} j_1^2 (k \tilde{R})\; e^{-\frac{1}{2} (k + \omega)^2} j_0(k \tilde{d}),\\
    \mathcal{Q} =& -\frac{\alpha}{2} e^{-\frac{1}{2} \omega ^2}\!\! \int_{0}^{\infty} \!\frac{\dd k}{ k}\; j_1^2 (k \tilde{R})\; w\!\left(\!\!-\frac{k}{\sqrt{2}}\right),\label{eq:Qcal_dimensionaless}\\
    \mathcal{M}_{ab}
    =& -\!\alpha e^{-\frac{1}{2} \omega ^2} \!\! \int_{0}^{\infty}\! \frac{\dd k}{ k} \;j_1^2 (k \tilde{R}) \; w\!\left(\!\!-\frac{k}{\sqrt{2}}\right) j_0(k \tilde{d}),\\
    \mathcal{V} =&\; \frac{\alpha}{2} e^{-\frac{1}{8} \omega ^2} \!\! \int_{0}^{\infty}\! \frac{\dd k}{k}\; j_1^2 (k \tilde{R}) \; w\!\left[\!\frac{1}{\sqrt{2}}\! \left(\frac{\omega }{2}-k \right)\!\right]\!,\label{eq:final_integrals}
\end{align}
where $\alpha= {9}\lambda^2/({4 \pi \tilde{R}^2})$, $\tilde{R}={R}/{T}$, $\omega=\Omega T$, $\tilde{d}={d}/{T}$, $j_0(z)$, $j_1 (z)$ are spherical Bessel functions, and $w(z)$ is the Faddeeva function.

\section{Contextual fraction for other fixed parameters}\label{app:more_plots}

\begin{figure}
    \centering
    \subfloat[]{%
    \includegraphics[width=0.48\linewidth]{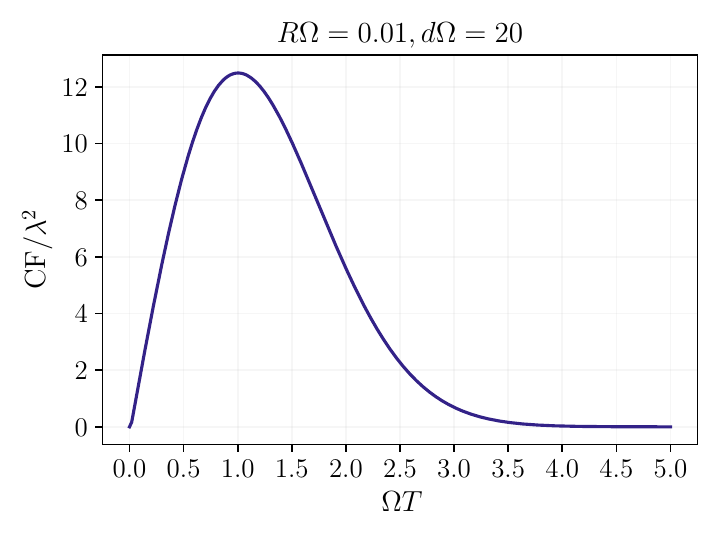}
    }\hfill
    \subfloat[]{
    \includegraphics[width=0.48\linewidth]{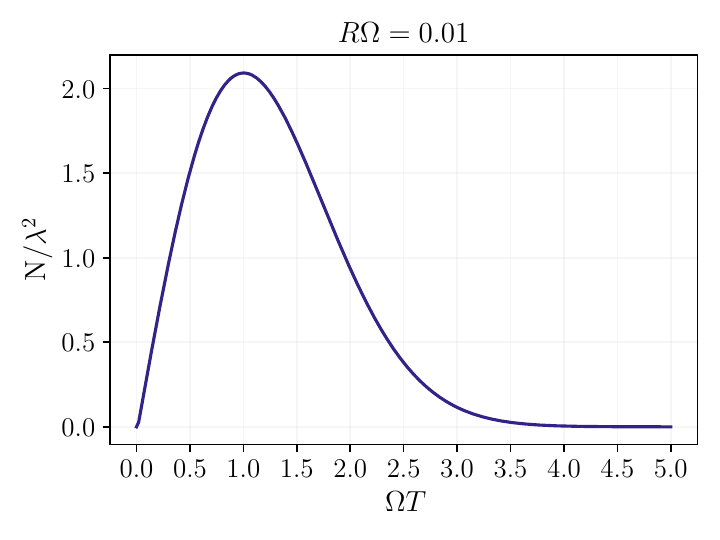}
    }
    \caption{Contextual fraction (a) and Wigner negativity (b) normalized by the squared coupling constant as a function of the product of the detectors' gap and their interaction duration. Here, instead of fixing $R/T$ and $d/T$, we fix $R\Omega$ and $d\Omega$.}
    \label{fig:comparisonmagic}
\end{figure}

In this appendix, we present the contextual fraction and Wigner negativity as functions of the dimensionless parameter $\Omega T$, with $R\Omega$ fixed, in contrast to \autoref{fig:mainplots}, where $R/T$ is fixed. We also provide more details on how our results agree with those from Ref.~\cite{Nystrom:2024oeq}.

With $R\Omega$ fixed (as in \autoref{fig:comparisonmagic}), contextual fraction and Wigner negativity vanish at $\Omega T=0$. However, with $R/T$ fixed (as in \autoref{fig:mainplots}), these quantities can be nonzero at $\Omega T=0$ and depend on the choice of $R/T$ in that limit. 
The difference becomes clear by examining the limits of $\mathcal{L}$ and $\mathcal{Q}$ as $\Omega T \to 0$. On the one hand, with $R/T$ fixed, the limits of $\mathcal{L}$ and $\mathcal{Q}$ depend on $R/T$. We discuss this fact in more detail for small $\tilde{R}=R/T$.
For small $\tilde{R}$,
\begin{align}
    \lim_{\omega \to 0}\mathcal{L} = \frac{1}{9} - \frac{2}{45} \tilde{R}^2 + \mathcal{O}(\tilde{R}^4), \quad \lim_{\omega \to 0}\operatorname{Re}\left[\mathcal{Q}\right] = -\frac{1}{2} \lim_{\omega \to 0}\mathcal{L}, \quad \lim_{\omega \to 0}\operatorname{Im}\left[\mathcal{Q}\right] = \frac{\sqrt{\pi/2}}{15 \tilde{R}} + \mathcal{O}(1).
\end{align}
Consequently, the inequality in Eq.~\eqref{eq:moreviolatedinequality} becomes, to leading order,
\begin{equation}
    \frac{3}{2} \left( \frac{1}{9} - \frac{2}{45}\tilde{R}^2\right) \ge \sqrt{3} \frac{\sqrt{\pi/2}}{15 \tilde{R}},
\end{equation}
which is violated for sufficiently small $\tilde{R}>0$. Hence, the Wigner negativity at $\Omega T \to 0$ can be non-zero and depends on $\tilde{R}=R/T$. 

On the other hand, we can rewrite Eqs.~\eqref{eq:Lcal_dimensionless} and \eqref{eq:Qcal_dimensionaless} using $R' = R\Omega$ (while retaining $\omega=\Omega T$):
\begin{align}
    \mathcal{L} = \frac{9\lambda^2}{4\pi} \frac{\omega^2}{R'^2} \int_{0}^\infty \frac{\dd k}{k} j_1^2\left( k R'\right)e^{-\frac{\omega^2}{2}\left( k+1\right)^2},\quad\mathcal{Q} = -\frac{9\lambda^2}{8\pi} \frac{\omega^2}{R'^2} \int_{0}^\infty \frac{\dd k}{k} j_1^2\left( k R'\right)w\left( -\frac{k\Omega}{\sqrt{2}} \right).
\end{align}
We see that, with $R'=R\Omega$ fixed, $\mathcal{L}$ and $\mathcal{Q}$ vanish as $\omega=\Omega T$ goes to 0. 
This limit is therefore independent of $R\Omega$, and the Wigner inequalities are trivially satisfied at $\Omega T=0$, rendering null contextual fraction and Wigner negativity at this point.  

Having clarified why CF and $N$ exhibit different qualitative behaviours depending on which parameters are kept fixed, we now use the fixed-$R\Omega$ version of our results to compare with Ref.~\cite{Nystrom:2024oeq}. This choice is natural because, near $\Omega T=0$, CF and Wigner negativity are independent of $R\Omega$; therefore, it  facilitates a comparison with the pointlike detectors $(R \to 0)$ considered in Ref.~\cite{Nystrom:2024oeq}.
Note that while Ref.~\cite{Nystrom:2024oeq} reports mana, we calculate Wigner negativity. However, the former is obtained as the logarithm of the latter so a qualitative comparison is still straightforward. Using \autoref{fig:comparisonmagic}, we indeed find qualitative agreement with Ref.~\cite{Nystrom:2024oeq}.

\end{widetext}

\end{document}